\documentclass[
	a4paper,
	10pt,
	twocolumn,
]{article}

\usepackage[
	top=2.5cm,
	bottom=2.5cm,
	left=2cm,
	right=2cm,
	footskip=1cm,
	headsep=0.75cm,
	columnsep=20pt,
]{geometry}

\usepackage{caption, subcaption, csquotes}

\usepackage{tikz, graphicx}
\usepackage{booktabs}
\usepackage{multirow}
\usepackage{braket, amsmath, amsthm, lipsum}

\usepackage{graphicx}
\usepackage[english]{babel}
\usepackage[utf8]{inputenc}
\usepackage[T1]{fontenc}

\usepackage{titlesec}
\titleformat
	{\section}
	[block]
	{\Large\bfseries\centering}
	{\thesection}
	{0.5em}
	{}
	[]

\usepackage{titling}
\pretitle{\begin{center}\huge\bfseries}
\posttitle{\end{center}}
\usepackage{abstract}

\usepackage{csquotes}
\usepackage[backend=bibtex,style=numeric,natbib=true,mcite=true,url=false]{biblatex}
\AtEveryBibitem{\clearfield{day}}
\AtEveryBibitem{\clearfield{month}}
\AtEveryBibitem{\clearfield{endday}}
\AtEveryBibitem{\clearfield{endmonth}}
\AtEveryBibitem{\clearfield{endyear}}
\AtEveryBibitem{\clearlist{location}}
\AtEveryBibitem{\ifentrytype{book}{}{\clearname{editor}}}

\title{\textbf{Implementation Of Dynamic De Bruijn Graphs\\Via Learned Index}}

\author{%
	Riccardo Nigrelli
}

\renewcommand{\maketitlehookd}{%
	\begin{abstract}
		De Bruijn graphs are essential for sequencing data analysis and must be
		efficiently constructed and stored for large-scale population studies. They
		also need to be dynamic to allow updates such as adding or removing edges
		and nodes. Existing dynamic implementations include \emph{DynamicBOSS} and
		dynamicDBG. In 2018, a new family of data structures called
		learned indexes was introduced by Tim Kraska and Alex
		Beutel, with a particularly efficient
		implementation proposed by Paolo Ferragina and Giorgio Vinciguerra in
		2020. This paper presents a new method for implementing De
		Bruijn graphs using learned indexes and compares its performance with
		current implementations. The new method shows improved time and memory
		efficiency for edge and node insertions, particularly with large 
		datasets (over 110 million k-mers).
	\end{abstract}
}

\begin{document}

\maketitle

\section{Introduction}
In the past two decades, genome assembly from vast amounts of DNA sequencing
data has been a major computational challenge in molecular biology, recognized
as an NP-hard problem~\cite{doi:10.1089/cmb.1995.2.275}. Traditionally,
long-read assembly algorithms use an overlap graph, where nodes represent reads
and edges exist if there is sufficient overlap between reads. This method,
however, struggles with the massive datasets produced by next-generation
sequencing (NGS) technologies due to computational constraints and the short
length of reads. 

To address these limitations, many recent algorithms have shifted to using de
Bruijn
graphs~\autocites{li2010novo}{Pevzner9748}{simpson2009abyss}{zerbino2008velvet}. 
In a de Bruijn graph~\cite{de1946combinatorial}, each node represents a $k$-mer
(a substring of length $k$ from the reads), and edges denote exact overlaps of
length $k-1$. Although de Bruijn graphs can be constructed more efficiently than
overlap graphs, they still require significant memory, making the overlap phase
a bottleneck.

Recent advances in combining data structures with Machine Learning (ML) have
introduced learned data structures~\cite{CaseOfLearnedIndex}, which replace
traditional components like arrays and trees with ML models. This approach
leverages data patterns to improve space efficiency and time performance across
various applications.

The aim of this article is to present a new implementation of de Bruijn graphs
using learned data structures, specifically starting with the
PGM-Index~\cite{PGMIndex}, an efficient learned index~\cite{Ferragina:2020icml}.
The performance, in terms of time and memory, is then compared with existing
dynamic implementations such as DynamicBOSS~\cite{SuccintDynamicDBG} and 
dynamicDBG~\cite{PracticalDynamicDBG}.

\section{Methodology}
Ferragina and Vinciguerra in~\cite{PGMIndex} not only present the PGM-Index from
a theoretical and computational perspective but also provide an implementation
in \verb!C++!. Therefore, it was decided to experimentally verify if this type
of learned index, specifically the dynamic version, would be a valid starting
point for the implementation of de Bruijn graphs. 

This particular implementation of a dynamic learned index is used when indexing
elements consisting of a key-value pair $(k, v)$ contained within a vector.
Indeed, the \emph{Dynamic PGM-Index} is implemented as a parametric class in $K$
and $V$, where $K$ and $V$ represent the types of the key and the value,
respectively. The behavior of the main operations are the following: 

\begin{itemize}
	\item creation: This involves creating $n$ empty levels up to the first level
	capable of containing all the elements of the input vector. A PGM-Index is
	then constructed and associated with this level.
	\item insertion: The correct insertion point within the first free level must
	be found so that the elements remain always sorted and the maximum capacity of
	the level is not violated. 
	\item deletion: The key-value pair is inserted into the lowest level capable
	of containing it, assigning the \emph{tombstone} to the value of the pair. 
	\item search: Using the key of the pair $(k,v)$, a level-by-level scan is
	performed starting from the lowest one. For each level, it is checked whether
	a PGM-Index is associated with it. If so, the search is performed within the
	PGM-Index; otherwise, it is conducted on the level itself. 
\end{itemize}

This type of learned index can be a good starting point for creating an
abstraction of de Bruijn graphs. However, the current implementation is very
limiting. Interacting only with sets whose elements are key-value pairs is
redundant for graphs since the value field would never be used. Therefore, it
was decided to improve the following aspects of the current implementation: the
ability to index vectors of single elements and not just key-value pairs,
finding a memory-efficient method for handling deletions, and implementing a
procedure to remove any duplicate and/or deleted elements. 

Additionally, it was decided to profile the heap memory during the creation
process to observe the trends and identify any possible points for further
optimization. It was found that at the end of the creation process, the memory
contains both the index and the vector on which it was built. To resolve this
issue, by utilizing the library offered by
KMC~\cite{DBLP:journals/bioinformatics/KokotDD17}, it was possible to develop an
\emph{online} index construction procedure. 

\section{Evaluation}
Once the implementation phase was completed, this new approach to implementing
de Bruijn graphs was verified and validated. Performance analysis in terms of
time and memory was conducted using the Dynamic PGM-Index Set on a real dataset,
\emph{E. Coli K-12 substr. MG1655}. Additionally, comparisons were made with existing
literature implementations: DynamicBOSS~\cite{SuccintDynamicDBG} and standard Dynamic PGM-Index. The E.
Coli dataset was chosen as it is commonly used by authors experimenting with
DynamicBOSS. Unfortunately, comparison with
dynamicDBG~\cite{PracticalDynamicDBG} was not possible as its runtime
implementation encountered errors with any input dataset. Four subsets were
generated from the initial dataset, comprising $20000$, $200000$, $2000000$ and
$14000000$ reads. 

For the creation operation, the complete dataset was used. For modification
operations (insertions/deletions) and searches, datasets were divided into two
parts: 80\% for creating data structures and 20\% for executing the operation.
Performance analysis for search operations was also conducted following
modifications to data structures. This involved inserting and/or deleting
elements from a graph followed by search operations to verify that modification
operations did not affect search operations. 
\begin{itemize}
	\item creation: The implementation based on sets shown better memory
	efficiency compared to the key-value pair-based approach, while having
	comparable time performance. DynamicBOSS, however, exhibited worse time
	performance across all datasets but proved the best solution in terms of
	memory (based solely on a BWT and bit vector). In
	Figure~\ref{fig:creation-result} are shown the results.
	\begin{figure}[!ht]
		\centering
		\subfloat[]{\label{fig:creation-time}\includegraphics[width=1\linewidth]{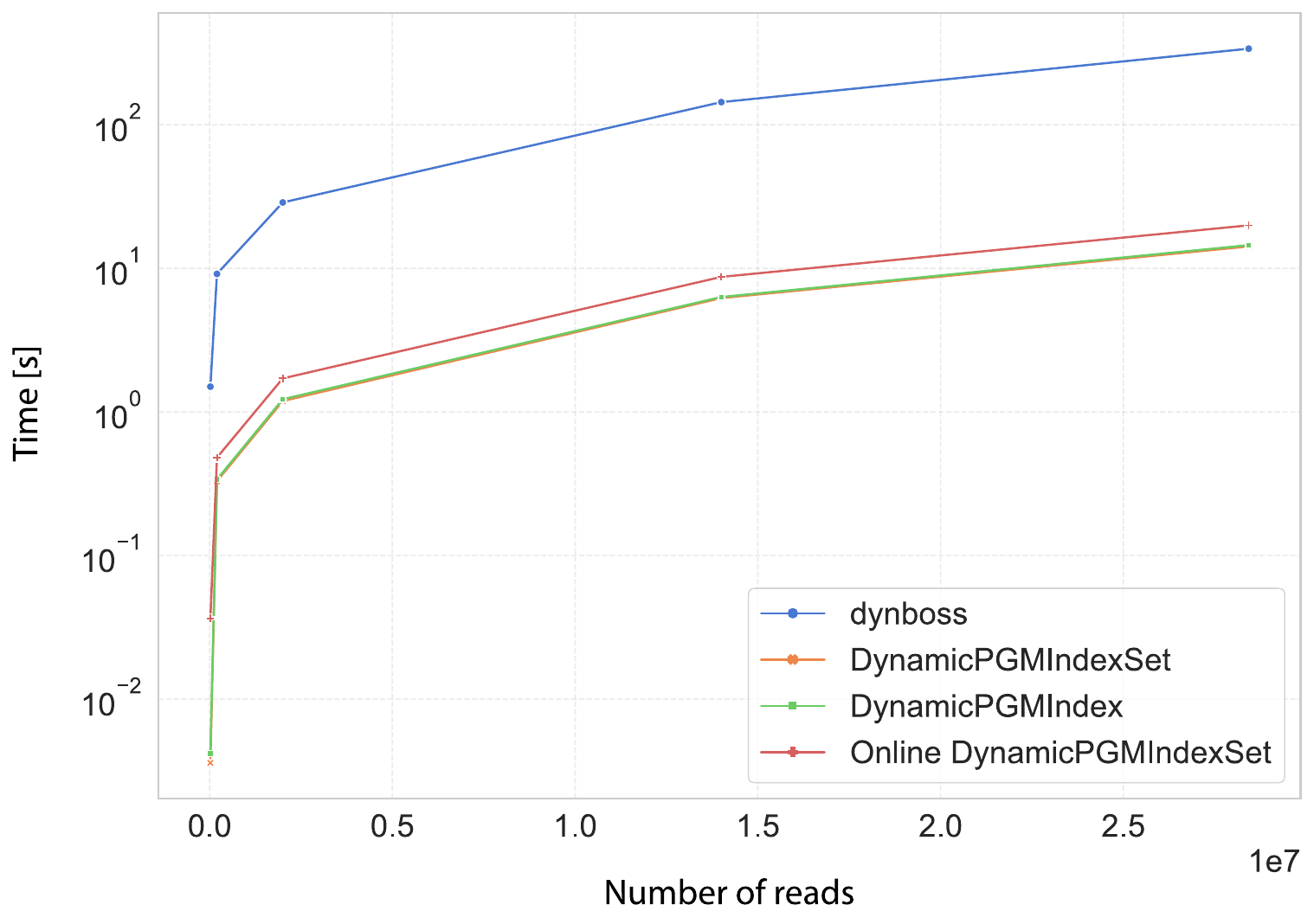}}\vspace{0.5cm}
		\subfloat[]{\label{fig:creation-memory}\includegraphics[width=1\linewidth]{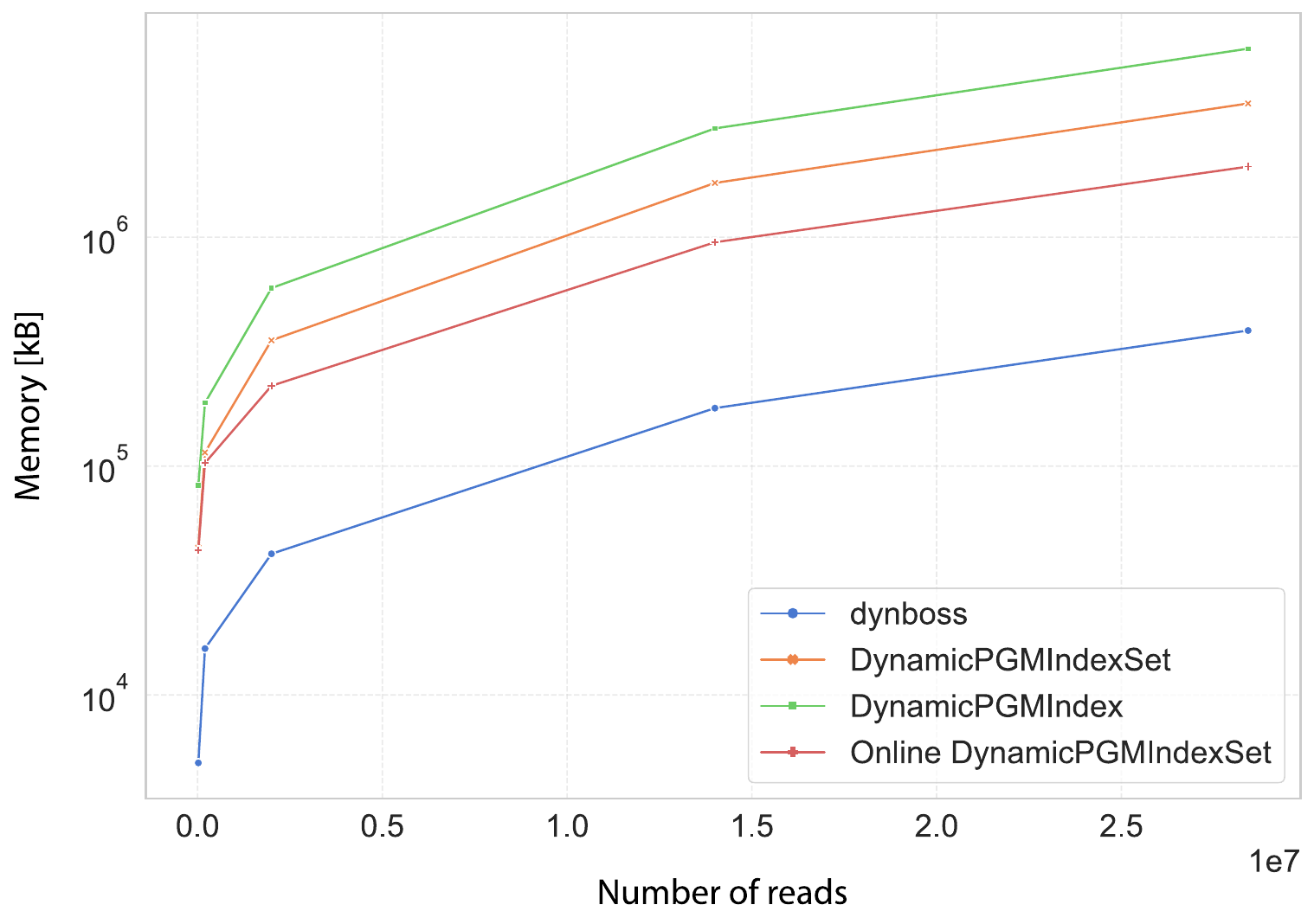}}
		\caption{Creation of a de Bruijn graph as a function of the number of reads.
		Note that the y-axis scale is logarithmic.~(\subref{fig:creation-time})
		Execution time~(\subref{fig:creation-memory}) Peak memory usage.}
		\label{fig:creation-result}
	\end{figure}
	
	\item insertion: DynamicBOSS could not be tested on larger datasets due to
	excessive execution time. Implementations based on PGM-Index, however, shown
	reasonable performance times across all datasets (just over half a minute on
	the largest dataset). In Figure~\ref{fig:insert-result} are shown the results.
	\begin{figure}[!ht]
		\centering
		\subfloat[]{\label{fig:insert-time}\includegraphics[width=1\linewidth]{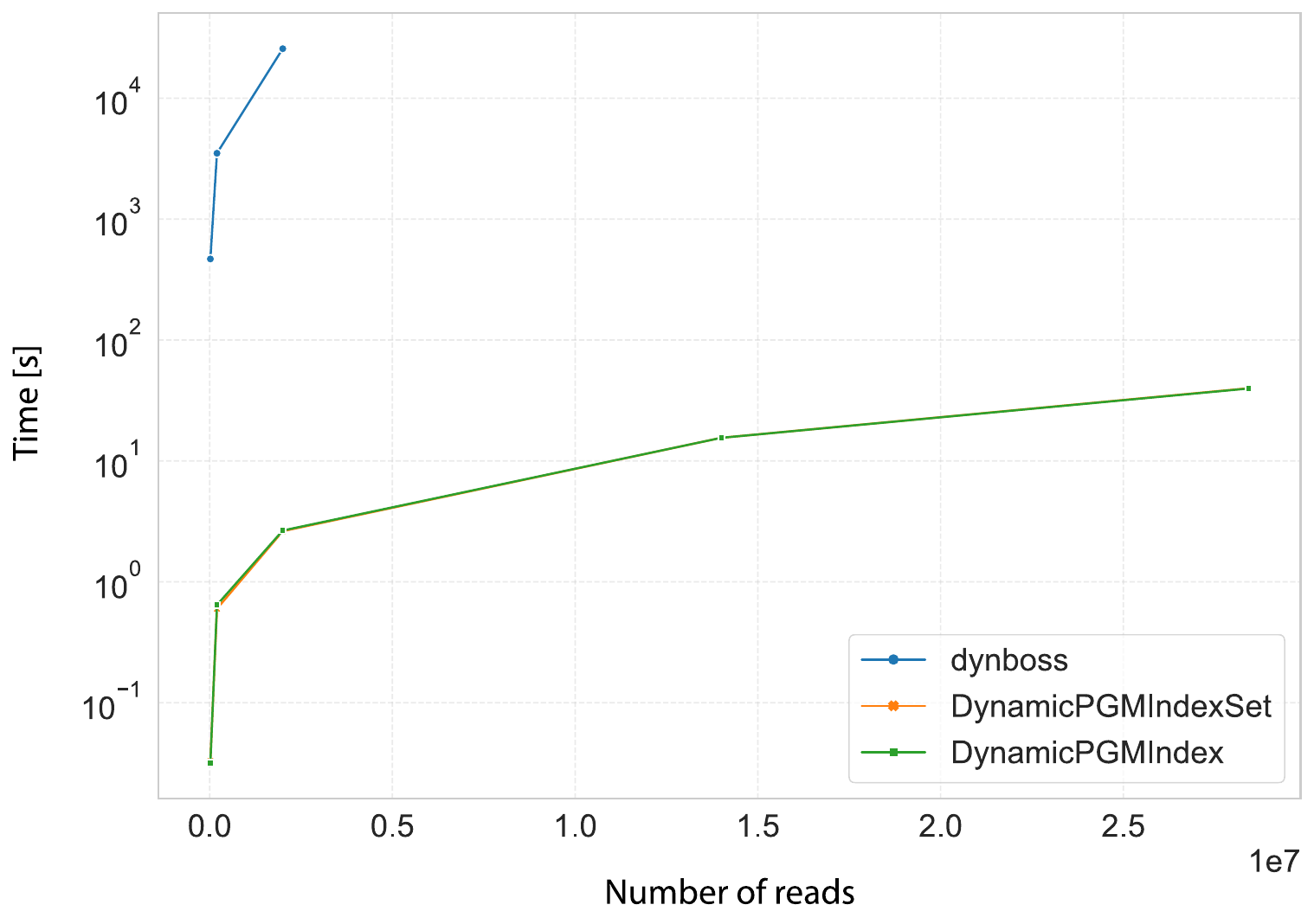}}\vspace{0.5cm}
		\subfloat[]{\label{fig:insert-memory}\includegraphics[width=1\linewidth]{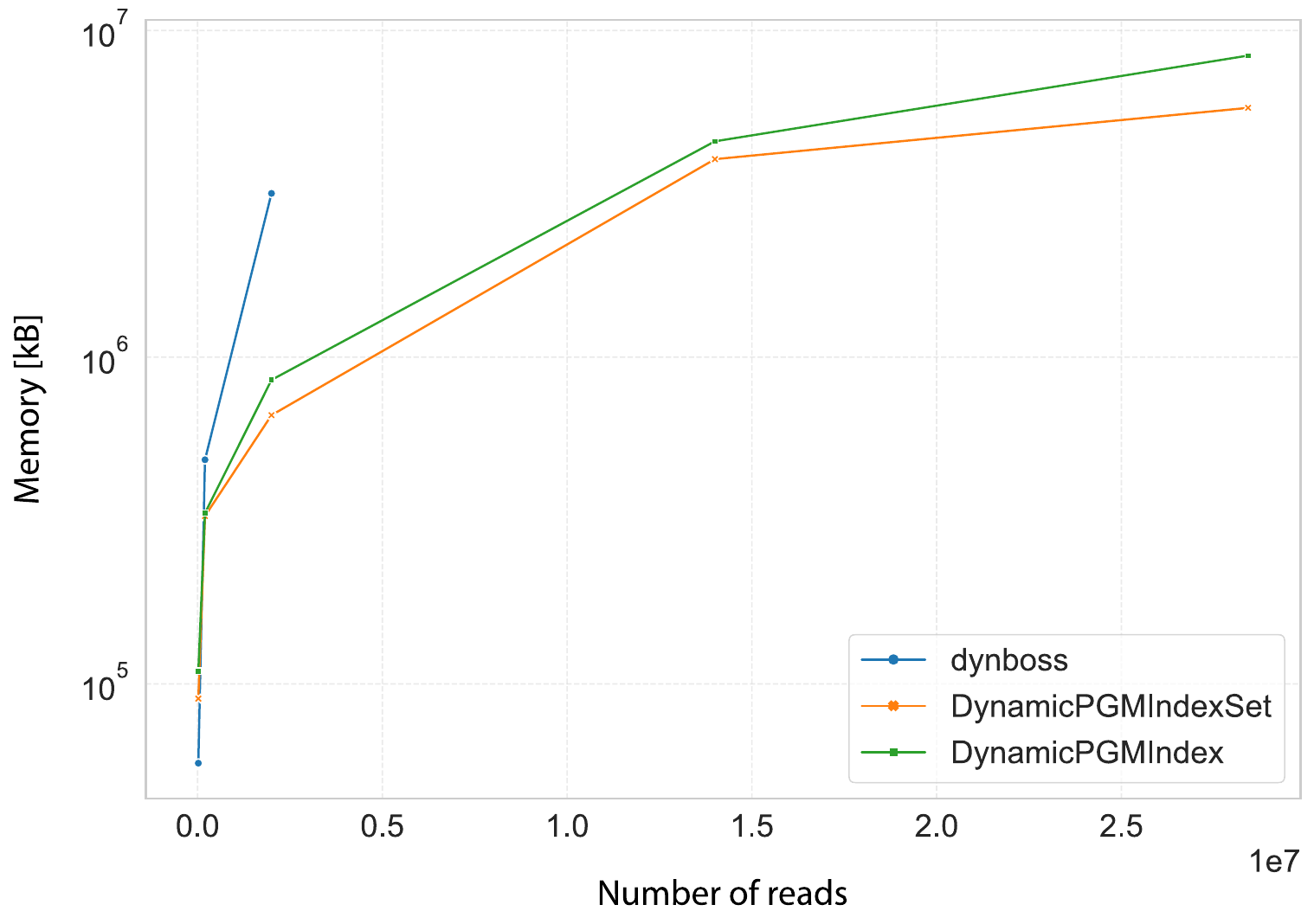}}
		\caption{Insertions within a de Bruijn graph as a function of the number of reads.
		Note that the y-axis scale is logarithmic.~(\subref{fig:insert-time})
		Execution time ~(\subref{fig:insert-memory}) Peak memory usage.}
		\label{fig:insert-result}
	\end{figure}

	\item deletion: Performance for deletion operations in PGM-Index-based
	implementations followed similar trends as insertion. Again, testing with
	DynamicBOSS could not be completed for the two largest datasets.
	In Figure~\ref{fig:delete-result} are shown the results.
	\begin{figure}[!ht]
		\centering
		\subfloat[]{\label{fig:delete-time}\includegraphics[width=1\linewidth]{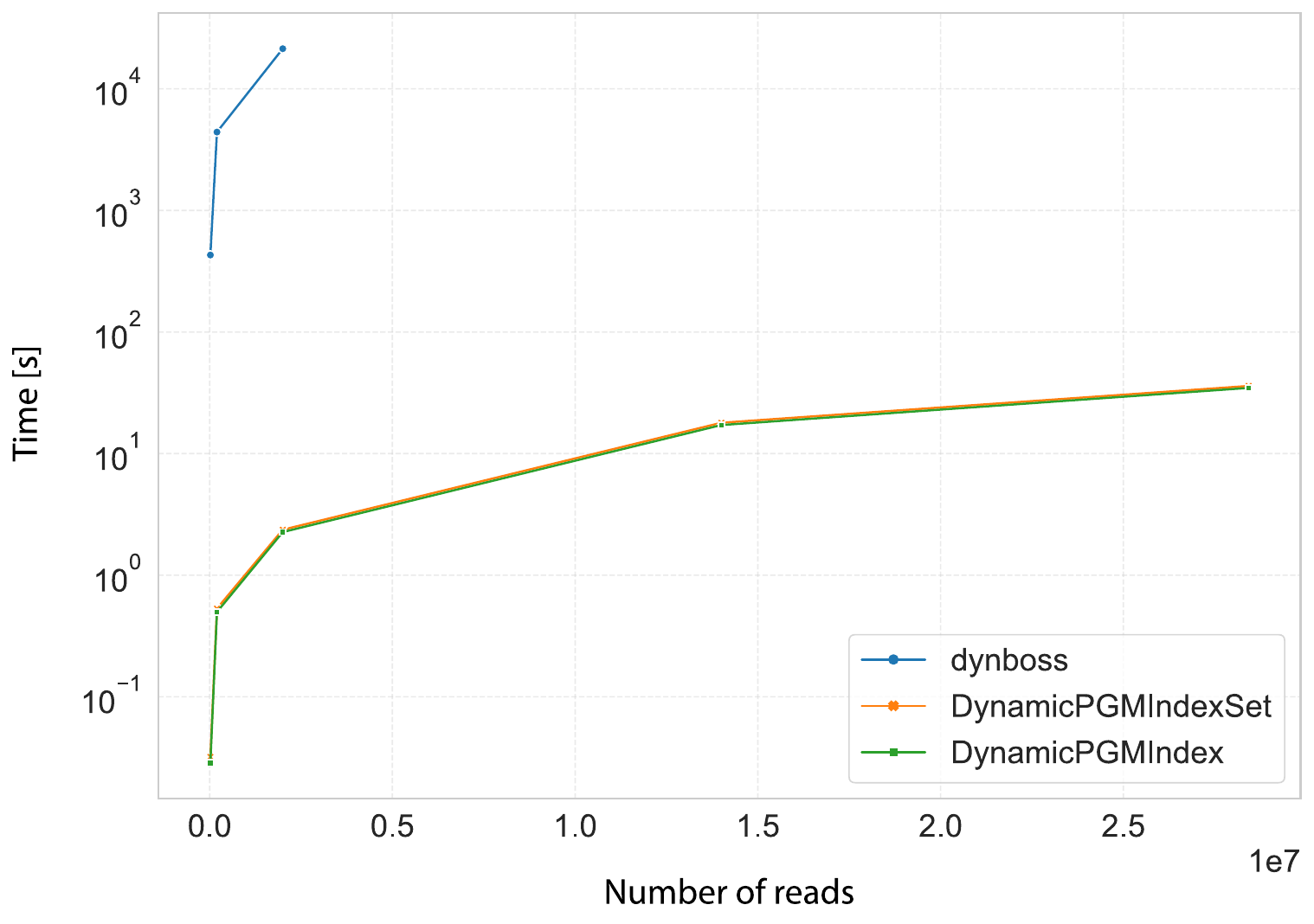}}\vspace{0.5cm}
		\subfloat[]{\label{fig:delete-memory}\includegraphics[width=1\linewidth]{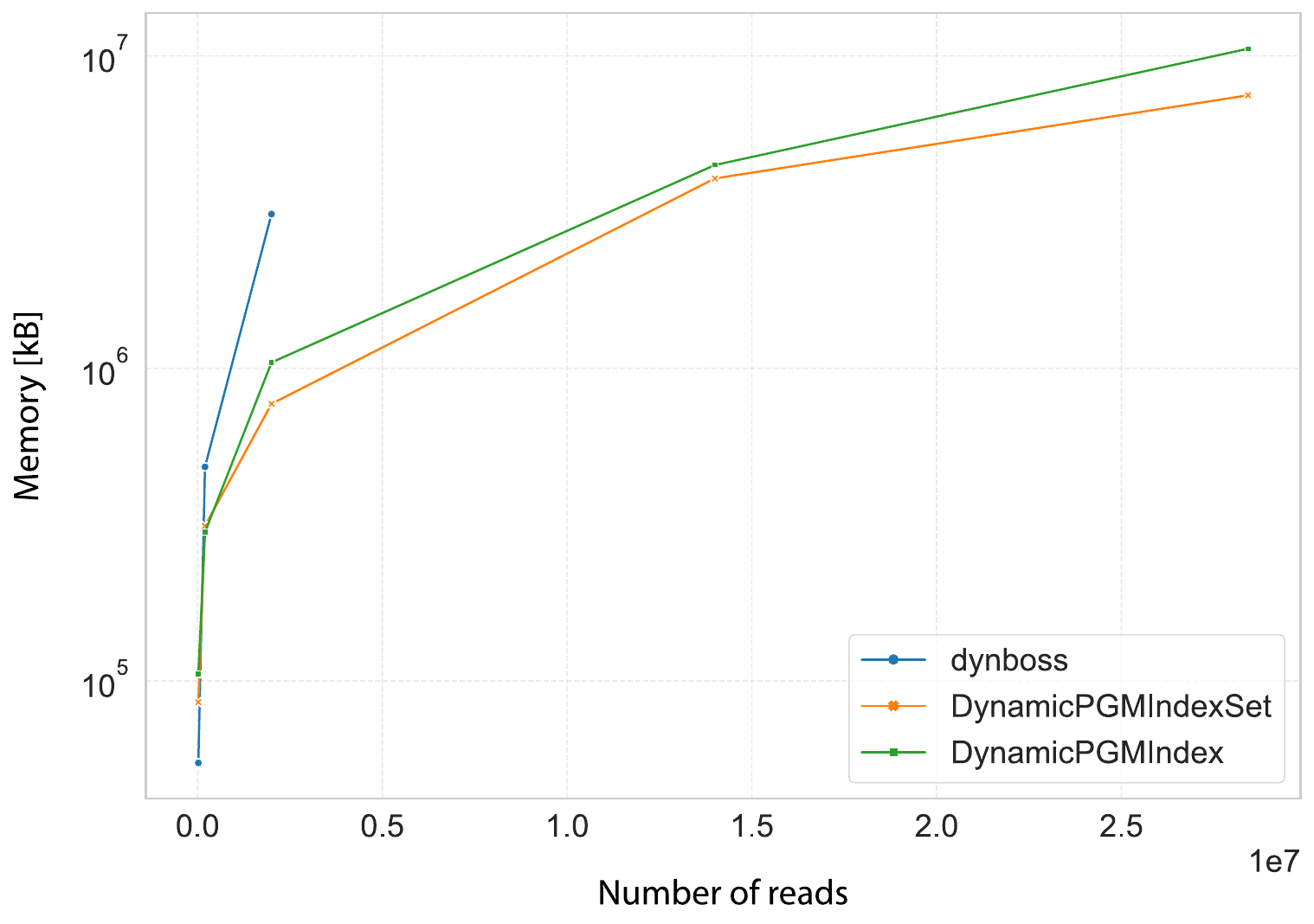}}
		\caption{Deletions within a de Bruijn graph as a function of the number of reads.
		Note that the y-axis scale is logarithmic.~(\subref{fig:delete-time})
		Execution time ~(\subref{fig:delete-memory}) Peak memory usage.}
		\label{fig:delete-result}
	\end{figure}

	\item search: DynamicBOSS did not demonstrate competitive performance in
	either time or memory. PGM-Index-based implementations shown equivalent time
	performance, with single-element-based approaches prevailing in terms of
	memory. Performance analysis of search operations following insertions and/or
	deletions yielded similar times and memory usage as those observed immediately
	after creation. In Figure~\ref{fig:search-result} are shown the results.
	\begin{figure}[!ht]
		\centering
		\subfloat[]{\label{fig:search-time}\includegraphics[width=1\linewidth]{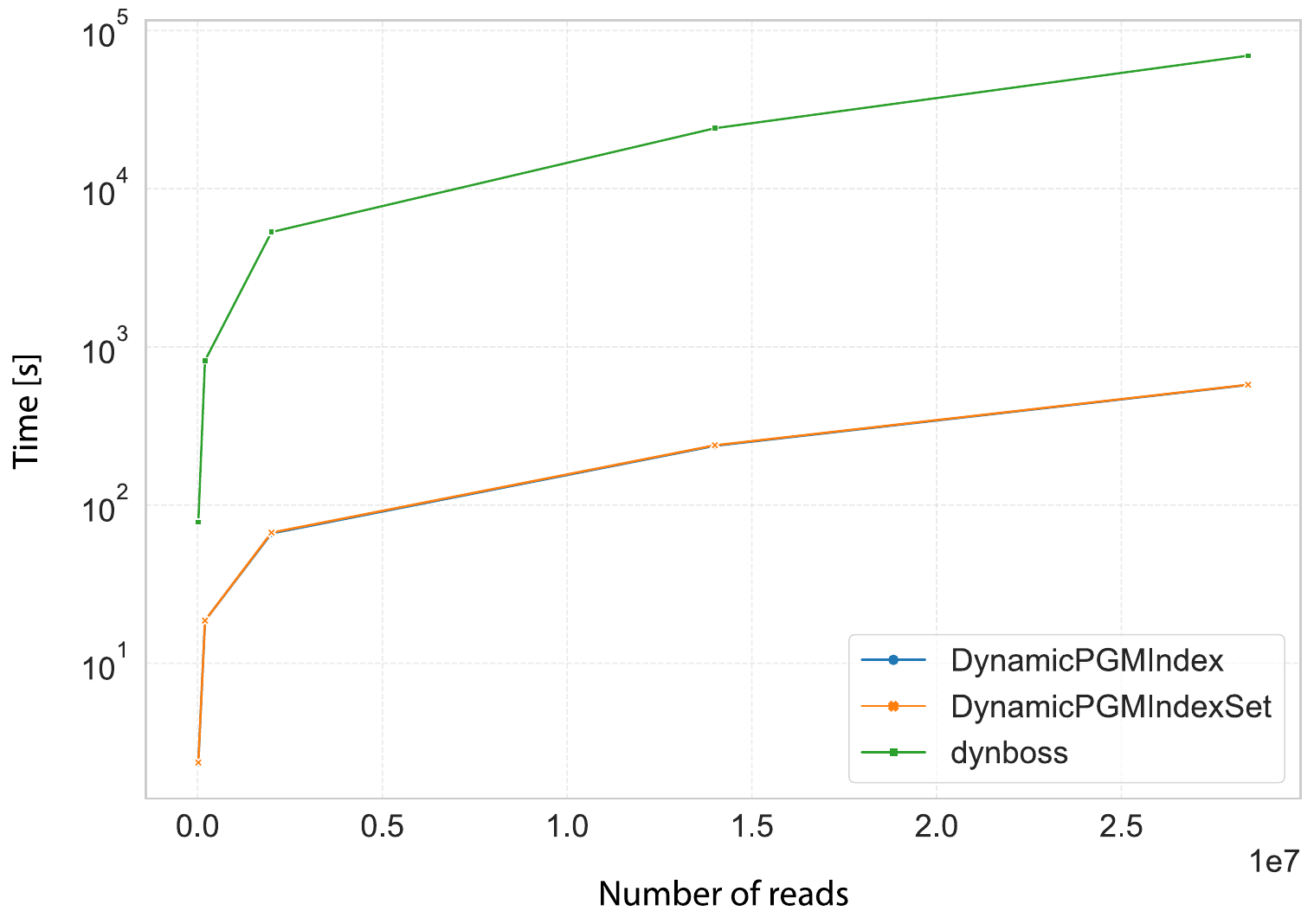}}\vspace{0.5cm}
		\subfloat[]{\label{fig:search-memory}\includegraphics[width=1\linewidth]{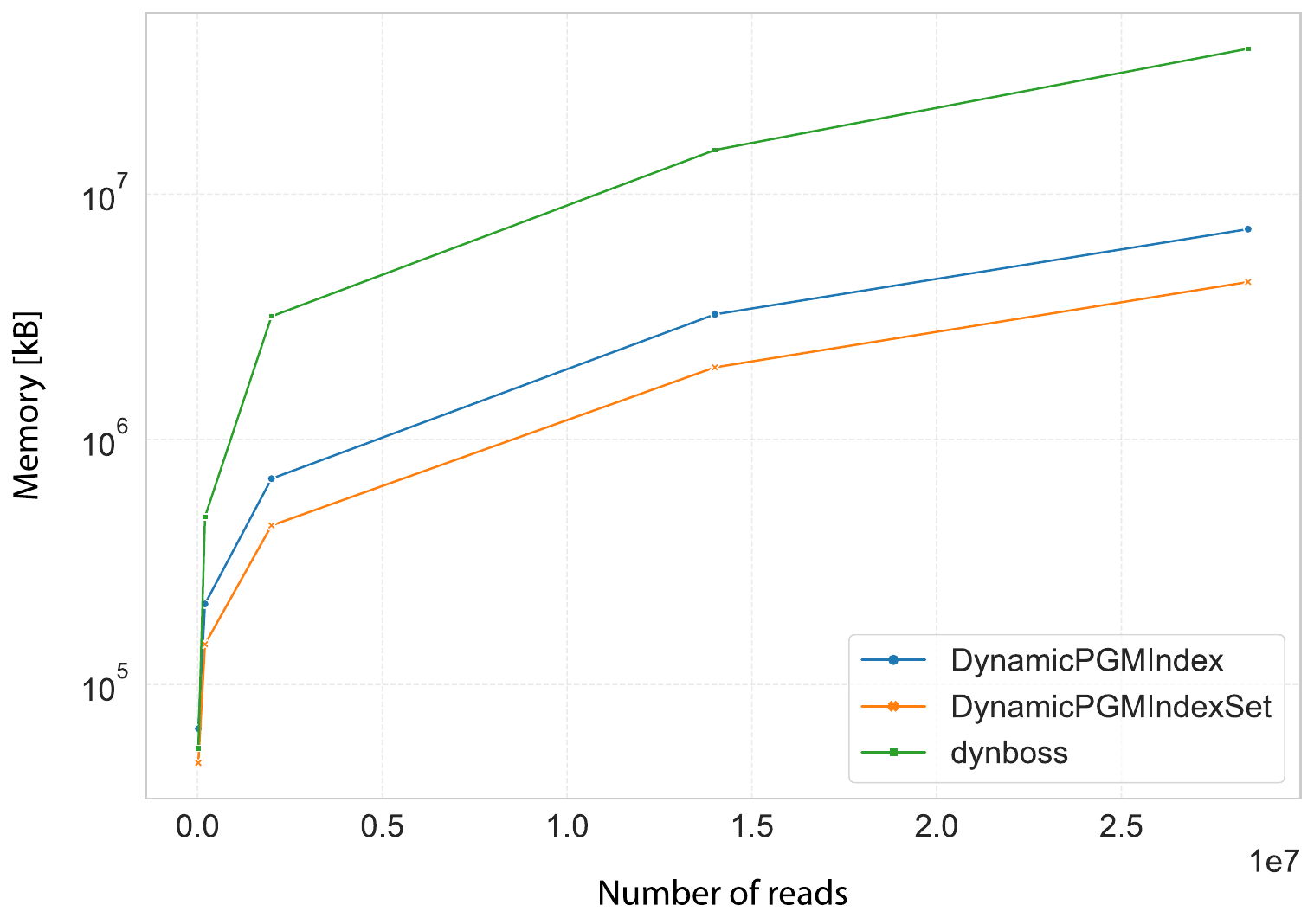}}
		\caption{Memberships within a de Bruijn graph as a function of the number of reads.
		Note that the y-axis scale is logarithmic.~(\subref{fig:search-time})
		Execution time ~(\subref{fig:search-memory}) Peak memory usage.}
		\label{fig:search-result}
	\end{figure}

\end{itemize}

\section{Conclusions and Future Works}
First and foremost, it should be highlighted that all the objectives set at the
beginning of the project have been achieved. The implementation of the Dynamic
PGM-Index Set data structure is derived from the implementation of the 
Dynamic PGM-Index, thus represented through a template to enable the indexing of
vectors of any numeric type. Moreover, all possible operations have been tested
through unit tests.

The implementation presented here proves to be a valid alternative to those
currently available in the literature for the representation of dynamic de
Bruijn graphs. In fact, it has been shown to be superior both in terms of time
and memory efficiency for both modification operations and searches. The
PGM-Index proved to be less efficient in terms of memory during the creation
process; however, despite this, it was still better in terms of time compared to
DynamicBOSS.

The current implementation of the \emph{Dynamic PGM-Index Set}
leaves points still open. Some future developments could be:
\begin{enumerate}
	\item extension of the representation of $k$-mers using the
	\emph{Wide-integer} library to allow for $k$ values up to 255, the maximum
	value in KMC. Currently, the implementation is limited to $k$-mers of length
	31. For the extension, a method needs to be found to convert the KMC
	representation (a vector of up to four elements) into the corresponding type
	offered by \emph{Wide-integer} for subsequent use in index construction; 
	\item implementation of a function to insert and/or delete a sorted set of
	$k$-mers without the need to iterate through it. One possible solution is to
	perform a \emph{pairwise\_merge} of the set with the first $i-1$ levels and use
	the result to build level $i$, where $i$ is the first level that has
	sufficient space to contain the result of the merge; 
	\item perform modifications and deletions of elements physically rather than
	logically. This could be implemented by locating the level containing the
	element to be modified or deleted, making the modification within the levels,
	and then reconstructing the associated PGM-Indexes. Another possibility is to
	clean the entire index to have a single level on which to perform the search,
	make the modification, and associate a new PGM-Index. 
\end{enumerate}
Another potential future development could be the feasibility analysis of using
\emph{Dynamic PGM-Indexes} for the representation of colored de Bruijn graphs.
In this case, indexing key-value pairs is useful because a specific color is
associated with each $k$-mer. If a $k$-mer can be associated with more than one
color, a bitmask could be used as the value, where each bit represents a color.

\printbibliography

@inproceedings{CaseOfLearnedIndex,
  author    = {Tim Kraska and
               Alex Beutel and
               Ed H. Chi and
               Jeffrey Dean and
               Neoklis Polyzotis},
  editor    = {Gautam Das and
               Christopher M. Jermaine and
               Philip A. Bernstein},
  title     = {The Case for Learned Index Structures},
  booktitle = {Proceedings of the 2018 International Conference on Management of
               Data, {SIGMOD} Conference 2018, Houston, TX, USA, June 10-15, 2018},
  pages     = {489--504},
  publisher = {{ACM}},
  year      = {2018}
}

@article{doi:10.1089/cmb.1995.2.275,
  author  = {Eugene W. Myers},
  title   = {Toward Simplifying and Accurately Formulating Fragment Assembly},
  journal = {J. Comput. Biol.},
  volume  = {2},
  number  = {2},
  pages   = {275--290},
  year    = {1995}
}

@inproceedings{Ferragina:2020icml,
  author    = {Paolo Ferragina and
               Fabrizio Lillo and
               Giorgio Vinciguerra},
  title     = {Why Are Learned Indexes So Effective?},
  booktitle = {Proceedings of the 37th International Conference on Machine Learning,
               {ICML} 2020, 13-18 July 2020, Virtual Event},
  series    = {Proceedings of Machine Learning Research},
  volume    = {119},
  pages     = {3123--3132},
  publisher = {{PMLR}},
  year      = {2020}
}

@article{Pevzner9748,
  author    = {Pevzner, Pavel A. and Tang, Haixu and Waterman, Michael S.},
  title     = {An Eulerian path approach to DNA fragment assembly},
  volume    = {98},
  number    = {17},
  pages     = {9748--9753},
  year      = {2001},
  publisher = {National Academy of Sciences},
  journal   = {Proceedings of the National Academy of Sciences}
}

@article{PGMIndex,
  author  = {Paolo Ferragina and
             Giorgio Vinciguerra},
  title   = {The PGM-index: a fully-dynamic compressed learned index with provable
             worst-case bounds},
  journal = {Proc. {VLDB} Endow.},
  volume  = {13},
  number  = {8},
  pages   = {1162--1175},
  year    = {2020}
}

@article{PracticalDynamicDBG,
  author  = {Victoria G. Crawford and
             Alan Kuhnle and
             Christina Boucher and
             Rayan Chikhi and
             Travis Gagie},
  title   = {Practical dynamic de Bruijn graphs},
  journal = {Bioinformatics},
  volume  = {34},
  number  = {24},
  pages   = {4189--4195},
  year    = {2018}
}

@article{SuccintDynamicDBG,
  doi       = {10.1093/bioinformatics/btaa546},
  url       = {https://doi.org/10.1093/bioinformatics/btaa546},
  year      = {2021},
  month     = jul,
  publisher = {Oxford University Press},
  volume    = {37},
  number    = {14},
  pages     = {1946--1952},
  author    = {Bahar Alipanahi and Alan Kuhnle and Simon J Puglisi and Leena Salmela and Christina Boucher},
  editor    = {Anthony Mathelier},
  title     = {Succinct dynamic de Bruijn graphs},
  journal   = {Bioinformatics}
}

@article{DBLP:journals/bioinformatics/KokotDD17,
  author    = {Marek Kokot and
               Maciej Dlugosz and
               Sebastian Deorowicz},
  title     = {{KMC} 3: counting and manipulating k-mer statistics},
  journal   = {Bioinform.},
  volume    = {33},
  number    = {17},
  pages     = {2759--2761},
  year      = {2017},
}

@inproceedings{de1946combinatorial,
  title={A combinatorial problem},
  author={de Bruijn, Nicolaas Govert},
  booktitle={Proc. Koninklijke Nederlandse Academie van Wetenschappen},
  volume={49},
  pages={758--764},
  year={1946}
}

@article{zerbino2008velvet,
  title={Velvet: algorithms for de novo short read assembly using de Bruijn graphs},
  author={Zerbino, Daniel R and Birney, Ewan},
  journal={Genome research},
  volume={18},
  number={5},
  pages={821--829},
  year={2008},
}

@article{simpson2009abyss,
  title={ABySS: a parallel assembler for short read sequence data},
  author={Simpson, Jared T and Wong, Kim and Jackman, Shaun D and Schein, Jacqueline E and Jones, Steven JM and Birol, Inan{\c{c}}},
  journal={Genome research},
  volume={19},
  number={6},
  pages={1117--1123},
  year={2009},
}

@article{li2010novo,
  title={De novo assembly of human genomes with massively parallel short read sequencing},
  author={Li, Ruiqiang and Zhu, Hongmei and Ruan, Jue and Qian, Wubin and Fang, Xiaodong and Shi, Zhongbin and Li, Yingrui and Li, Shengting and Shan, Gao and Kristiansen, Karsten and others},
  journal={Genome research},
  volume={20},
  number={2},
  pages={265--272},
  year={2010},
}

\end{document}